\documentclass[lettersize,journal]{IEEEtran}
\usepackage{amsmath,amsfonts}
\usepackage{physics,amsmath}
\usepackage{algorithm}
\usepackage{algpseudocode}
\usepackage{multirow}
\usepackage{array}
\usepackage{bm}
\usepackage{placeins}
\usepackage{amsthm}
\usepackage{amsmath,amsthm}
\usepackage{amsthm} 
\newtheorem{lemma}{Lemma}  

\newtheorem{proposition}{Proposition} 
\usepackage {wrapfig}
\usepackage{xcolor}
\usepackage{tcolorbox}
\usepackage{amsfonts}

\usepackage{amsmath,amssymb}

\usepackage{subcaption}
\usepackage{mathtools}
\usepackage{xcolor}
\usepackage{float} 
\usepackage{textcomp}
\usepackage{stfloats}
\usepackage{amsmath}   

\usepackage{lipsum}  
\usepackage{verbatim}
\usepackage{graphicx}
\usepackage{pdfpages}
\usepackage{tikz}
\usepackage{lettrine}
\usetikzlibrary{3d}
\usetikzlibrary{arrows.meta} 

\usepackage[colorlinks=true, linkcolor=blue, urlcolor=blue, citecolor=blue, anchorcolor=blue]{hyperref}

\begin{document}

\title{Fundamental Limits of THz Inter-Satellite ISAC Under Hardware Impairments}
\author{Haofan Dong,~\IEEEmembership{Student Member,~IEEE},
        and Ozgur B. Akan,~\IEEEmembership{Fellow,~IEEE}
\thanks{The authors are with Internet of Everything Group, Department of Engineering, University of Cambridge, CB3 0FA Cambridge, UK.}
\thanks{Ozgur B. Akan is also with the Center for neXt-generation Communications
(CXC), Department of Electrical and Electronics Engineering, Koç University, 34450 Istanbul, Turkey (email:oba21@cam.ac.uk)}}
	   



\maketitle

\begin{abstract}
This paper establishes a theoretical framework for analyzing the fundamental performance limits of terahertz (THz) Low Earth Orbit (LEO) inter-satellite link (ISL) Integrated Sensing and Communications (ISAC) systems. We develop a unified, end-to-end signal model that, jointly captures the effects of extreme orbital dynamics, cascaded non-ideal hardware impairments, and micro-radian beam pointing errors. Through Bayesian Cramér-Rao Lower Bound (BCRLB) analysis, we derive the ultimate sensing accuracy for range and range-rate, revealing a quadratic ($1/f_c^2$) improvement in estimation variance with carrier frequency, which is ultimately floored by signal-dependent hardware distortion. For communication, we show that system performance is not power-limited but hardware-limited, deriving a closed-form capacity ceiling under the joint effect of phase noise and PA nonlinearity: $C_{\text{sat}} = \log_2(1 + e^{-\sigma_\phi^2}/\Gamma_{\text{eff}})$, where $\Gamma_{\text{eff}}$ is a proposed hardware quality factor. Our numerical results, based on state-of-the-art component data and the identified trade-offs, suggest that favorable operational conditions may exist in the sub-THz frequency range (200-600 GHz) where the quadratic sensing gain with frequency is balanced against hardware quality degradation. Power Amplifier (PA) nonlinearity emerges as the dominant performance bottleneck, exceeding other impairments by one to two orders of magnitude.
\end{abstract}

\begin{IEEEkeywords}
ISAC, THz communications, LEO satellites, capacity-distortion trade-off, Cramér-Rao lower bound, hardware impairments.
\end{IEEEkeywords}

\section{Introduction}

\IEEEPARstart{T}{he} proliferation of Low Earth Orbit (LEO) mega-constellations is creating a demand for multi-Tbps Inter-Satellite Links (ISLs) that legacy Ka/V-band technologies cannot meet \cite{ullah2025beyond, pi2022dynamic, chen2025review}. The terahertz (THz) band (0.1-1 THz) offers a solution, providing vast bandwidths ($>$100 GHz) and enabling highly directional beams suitable for secure, high-capacity ISLs \cite{memioglu2023300, marcus2021spectrum}. Beyond raw data rate, Integrated Sensing and Communications (ISAC) functionalities into these THz links promises revolutionary capabilities for 6G-era constellations. These include sub-millimeter relative positioning \cite{wang2025fundamental}, real-time debris detection and classification as demonstrated in DebriSense \cite{dong2025debrisense}, and potential extensions to deep space networks for environmental monitoring \cite{dong2025masc, dong2024martian}. However, realizing these synergistic benefits in the extreme LEO-ISL environment—characterized by a complex deterministic-randomness tradeoff (DRT), extreme kinematics, and stringent pointing requirements—necessitates a rigorous understanding of the underlying fundamental performance limits \cite{liu2022integrated, mu2023uav}.

However, existing theoretical frameworks, developed primarily for terrestrial millimeter-wave (mmWave) systems, fundamentally fail to capture the unique confluence of challenges inherent to THz LEO-ISL ISAC. First, the extreme orbital kinematics create unprecedented signal distortions: relative velocities reaching 15 km/s generate Doppler shifts exceeding 50 MHz at 1 THz \cite{marchetti2021space}, while the differential Doppler across ultra-wide bandwidths manifests as a severe Doppler Squint Effect (DSE) that can reach 5 MHz across the band \cite{you2022beam}. Second, unlike power-limited terrestrial systems, THz transceivers operate in a fundamentally hardware-limited regime where cascaded impairments—phase noise degrading as $20\log_{10}(f)$ with frequency multiplication \cite{le2023performance}, and Power Amplifier (PA) nonlinearities creating signal-dependent distortions \cite{chen2025review}—impose hard performance ceilings. Third, the ISL's line-of-sight vacuum channel eliminates multipath fading but introduces an equally challenging pointing-error-dominated regime: with beamwidths narrowing to 2-4 milliradians at 300 GHz (corresponding to an antenna diameter of approximately 0.25-0.5~m) \cite{khan2011mmwave}, even microradian-level platform vibrations cause dramatic gain fluctuations \cite{sidhu2021advances}. Critically, no existing theoretical framework simultaneously captures these three interdependent phenomena within a unified model.

To address this gap, this paper develops a unified theoretical framework that, jointly models the extreme kinematics, cascaded hardware impairments, and beam pointing errors in THz LEO-ISL ISAC systems. Our contributions and key findings are as follows: \textit{(i)} We establish a comprehensive signal model integrating orbital dynamics, DSE, signal-dependent noise via Bussgang decomposition, and pointing-induced fading. \textit{(ii)} We derive the Bayesian Cramér-Rao Lower Bound (BCRLB) for joint kinematic estimation, revealing how hardware impairments fundamentally couple with sensing accuracy. \textit{(iii)} We derive a novel, closed-form capacity ceiling, $C_{\text{sat}} = \log_2(1 + e^{-\sigma_\phi^2}/\Gamma_{\text{eff}})$, proving that system performance is hardware-limited, not power-limited. \textit{(iv)} Through this framework, our analysis reveals that PA nonlinearity is the dominant performance bottleneck by one to two orders of magnitude ($\Gamma_{\text{PA}}/\Gamma_{\text{LO}} \approx 10^4$), suggests an optimal operational frequency window of 200-600 GHz, and quantifies a 3.38 bits/symbol capacity gain from improving the hardware quality factor $\Gamma_{\text{eff}}$ from 0.05 to 0.005, with State-of-the-Art hardware achieving 7.56 bits/symbol ceiling compared to 4.18 bits/symbol for Low-Cost solutions.

The remainder of this paper is organized as follows: Section II develops the unified system model; Section III derives the performance analysis including both sensing BCRLB and communication capacity limits; Section IV presents numerical results validating the theoretical framework; and Section V concludes with design principles and future research directions.

\section{System Model}

\begin{figure}[!t]
    \centering
    \includegraphics[width=\columnwidth]{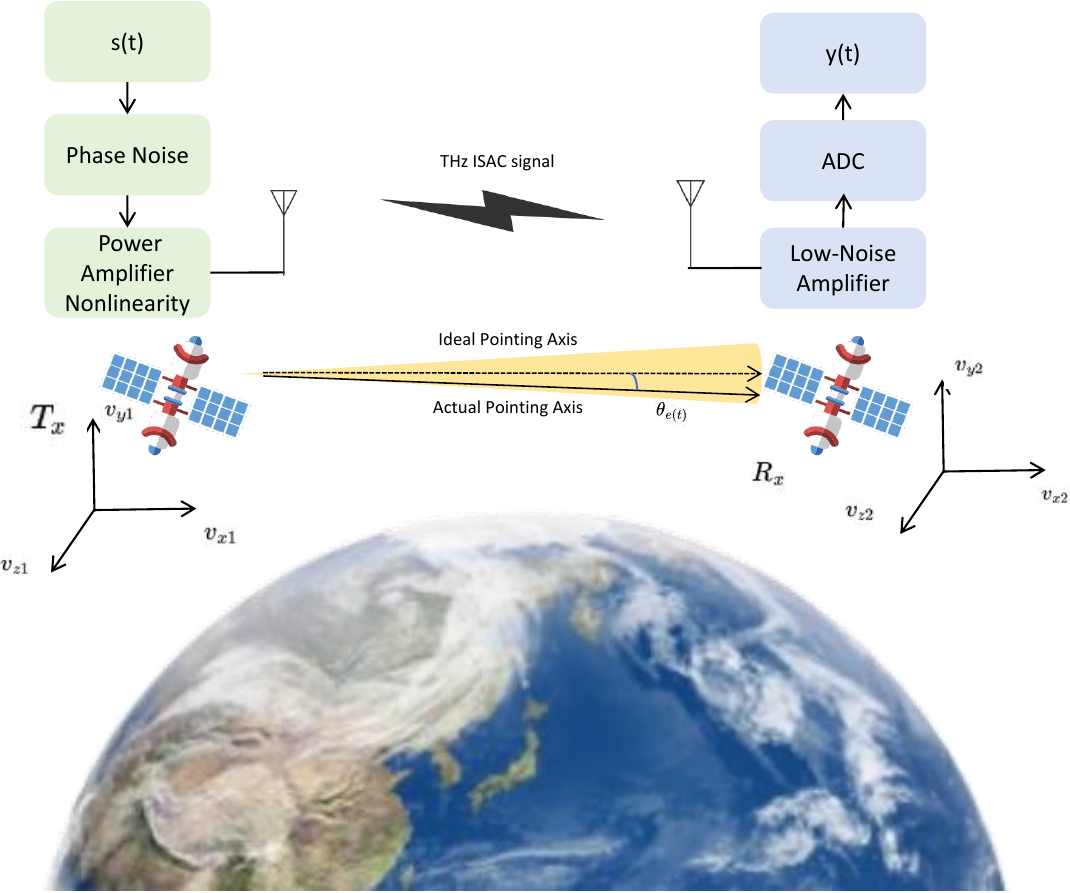}
    \caption{System model for the THz LEO-ISL ISAC scenario, illustrating the key challenges.}
    \label{fig:system_model}
\end{figure}

We consider a THz ISAC system for LEO ISLs, as illustrated in Fig.~\ref{fig:system_model}. The system operates at carrier frequencies $f_c \in [100, 600]$ GHz, where large bandwidth availability enables Tbps-level data rates while narrow beamwidths provide precise sensing capabilities. LEO satellites maintain relative distances of $R \in [500, 5000]$ km with relative velocities up to $v_{\text{rel}} = 15$ km/s \cite{reus2024simulation}, creating extreme Doppler dynamics absent in terrestrial systems. Our analysis framework is built upon a unified model that captures these macroscopic orbital dynamics and microscopic hardware-level impairments.

\subsection{Signal Structure and Parameters}

The ISAC waveform employs $M$ orthogonal pilot symbols for joint parameter estimation and data transmission. The baseband signal vector $\mathbf{x} \in \mathbb{C}^{M \times 1}$ undergoes digital precoding:
\begin{equation}
\mathbf{s}_{\text{BB}} = \mathbf{W}_{\text{BB}} \mathbf{x},
\label{eq:baseband_signal}
\end{equation}
where $\mathbf{W}_{\text{BB}} \in \mathbb{C}^{N_t \times M}$ is the baseband precoding matrix designed to optimize the sensing-communication trade-off.

\textbf{Observable Parameter Limitations:} For a single ISL, the observable parameter space is fundamentally limited by the one-dimensional nature of the propagation path. While the complete satellite kinematic state requires an 8-dimensional parameter vector $\boldsymbol{\eta}_{\text{full}} = [\Delta\mathbf{R}^T, \Delta\mathbf{V}^T, \boldsymbol{\theta}_e^T]^T \in \mathbb{R}^8$ (comprising 3D position error $\Delta\mathbf{R}$, velocity error $\Delta\mathbf{V}$, and 2D pointing error $\boldsymbol{\theta}_e$), a single ISL can only observe projections along the line-of-sight direction. Specifically, we can estimate:
\begin{itemize}
    \item Range: $R = \|\Delta\mathbf{R}\|$ (scalar distance)
    \item Range-rate: $\dot{R} = \Delta\mathbf{V} \cdot \hat{\mathbf{u}}_{\text{LOS}}$ (radial velocity component)
\end{itemize}
The complete 3D state observability requires multiple non-coplanar ISLs, analogous to trilateration principles in Global Navigation Satellite Systems (GNSS). Therefore, our BCRLB analysis focuses on these two observable parameters, with the understanding that network-level fusion would enable full state estimation.

While the proposed signal model captures the dominant hardware impairments, it is acknowledged that other non-idealities, such as I/Q imbalance (IQI), are present in practical THz transceivers. IQI introduces a mirror-frequency interference term proportional to the conjugated signal, typically modeled as \cite{reus2024simulation}:
\begin{equation}
s_{\text{IQI}}(t) = K_1 s(t) + K_2 s^*(t),
\end{equation}
where $K_1$ and $K_2$ represent the imbalance coefficients.

However, the model assumes IQI has been compensated to a residual level considered negligible through standard digital pre-distortion, allowing us to focus on the more fundamental phase noise and PA nonlinearity limitations.

\subsection{End-to-End Signal Model}

The received signal undergoes multiple transformations through the ISL channel. We model the mean received signal as:
\begin{equation}
\boldsymbol{\mu}_y = \Pi(\boldsymbol{\theta}_e) \cdot H_{\text{dyn}}(t,f;\boldsymbol{\eta}) \cdot f_{\text{PA}}(s_{\text{in}}).
\label{eq:rx_signal_mean_full}
\end{equation}

Each component captures distinct physical phenomena that we now detail.

\subsubsection{Beam Pointing Loss Model}

THz systems employ highly directive antennas to overcome severe path loss. For a typical 0.5m antenna at 300 GHz, the beamwidth is \cite{willey2002antenna}:
\begin{equation}
\theta_{3\text{dB}} = 1.02 \frac{\lambda}{D} \approx 0.12^\circ= 2.1 \text{ mrad}.
\end{equation}

Such narrow beams create extreme sensitivity to pointing errors. Using a Gaussian beam pattern approximation valid for $|\boldsymbol{\theta}_e| < 2\theta_{3\text{dB}}$ \cite{valentini2021analysis}:
\begin{equation}
\Pi(\boldsymbol{\theta}_e) = \exp\left(-g_{\text{power}}\|\boldsymbol{\theta}_e\|^2\right),
\label{eq:pointing_loss_full}
\end{equation}
where $\boldsymbol{\theta}_e = [\theta_x, \theta_y]^T$ represents the 2D angular misalignment in the antenna frame, and $g_{\text{power}} = 4\ln(2)/\theta_{3\text{dB}}^2 \approx 2.77/\theta_{3\text{dB}}^2$ is the power rolloff factor for a Gaussian beam pattern. Since the channel gain $g(\boldsymbol{\eta})$ represents the complex amplitude, the pointing-induced amplitude attenuation is $\sqrt{\Pi(\boldsymbol{\theta}_e)} = \exp(-g_{\text{power}}\|\boldsymbol{\theta}_e\|^2/2)$. Therefore, in gradient calculations, we use $\gamma = g_{\text{power}}/2 = 2\ln(2)/\theta_{3\text{dB}}^2$.

\subsubsection{Dynamic Channel Response}

The propagation channel for ISL exhibits unique characteristics:
\begin{equation}
H_{\text{dyn}}(t,f;\boldsymbol{\eta}) = |g(\boldsymbol{\eta})| \cdot e^{j\Phi_{\text{carrier}}(t;\boldsymbol{\eta})} \cdot e^{j\Phi_{\text{DSE}}(t,f;\boldsymbol{\eta})}
\label{eq:channel_dynamic_full}
\end{equation}

The complex channel gain $g(\boldsymbol{\eta})$ has magnitude following the Friis equation:
\begin{equation}
|g(\boldsymbol{\eta})| = \frac{c}{4\pi R f_c}\sqrt{G_{\text{tx}}G_{\text{rx}}}.
\end{equation}
We denote the complex channel gain as $g(\boldsymbol{\eta})$ throughout this paper.

The carrier phase enables precise ranging:
\begin{equation}
\Phi_{\text{carrier}} = -\frac{2\pi f_c R}{c}.
\end{equation}
A 1 mm range change causes a phase shift of $2\pi$ at 300 GHz, enabling sub-millimeter ranging accuracy.

The differential Doppler stretch effect (DSE) arises from the frequency-dependent Doppler across the wide THz bandwidth \cite{wang2023doppler}. For a 100 GHz bandwidth and 15 km/s relative velocity, the differential Doppler across the band reaches 5 MHz, which would cause catastrophic inter-symbol interference if left uncompensated. This framework assumes that DSE compensation is performed at the receiver using established techniques such as OTFS modulation or sub-band processing. The residual DSE after compensation is modeled as:
\begin{equation}
\Phi_{\text{DSE}}(t,f;\boldsymbol{\eta}) = 2\pi \left( f_D \frac{f}{f_c} t + \frac{1}{2} \dot{f}_D \frac{f}{f_c} t^2 \right),
\label{eq:dse_phase_full}
\end{equation}
where $f$ is the baseband frequency offset, $f_D = -f_c v_{\text{rel}}/c$ is the Doppler shift, and $\dot{f}_D = -f_c a_{\text{rel}}/c$ is the Doppler rate. The residual uncompensated DSE power is incorporated as $\sigma^2_{\text{DSE}}$ in the effective noise variance.

The differential Doppler effect arises from the frequency-dependent phase evolution. For a signal at baseband frequency $f$:
\begin{align}
\phi(t,f) &= 2\pi(f_c + f)\frac{R(t)}{c} \nonumber\\
&= 2\pi f_c \frac{R_0 - v_{\text{rel}}t}{c} + 2\pi f \frac{R_0 - v_{\text{rel}}t}{c} \nonumber\\
&= \phi_{\text{carrier}}(t) + 2\pi f \left(\frac{R_0}{c} - \frac{v_{\text{rel}}t}{c}\right).
\end{align}

Taking the time derivative gives the instantaneous frequency. The total instantaneous frequency is scaled by the Doppler effect:
\begin{equation}
    f'_{\text{total}}(t,f) = (f_c+f) \left(1 - \frac{v_{\text{rel}}}{c}\right) = (f_c+f) \left(1 + \frac{f_D}{f_c}\right)
    \label{eq:inst_freq_corrected}
\end{equation}
This frequency scaling across the band creates the DSE phase term in Eq.~\eqref{eq:dse_phase_full}.

\subsubsection{Transmitter Hardware Impairments}

The signal entering the PA incorporates multiple impairments:
\begin{equation}
s_{\text{in}} = e^{j\phi_{\text{PN,eff}}(t)} \cdot \mathcal{T}_{\text{IQI}}\{\mathbf{s}_{\text{BB}}\}.
\label{eq:pa_input_full}
\end{equation}

The effective phase noise $\phi_{\text{PN,eff}}(t)$ combines local oscillator phase noise and digital-to-analog converter (DAC) clock jitter:
\begin{equation}
\mathcal{L}_{\phi,\text{eff}}(f_m) = \mathcal{L}_{\text{LO}}(f_m) + (2\pi f_m \sigma_{t,\text{DAC}})^2,
\end{equation}
where $\mathcal{L}(f_m)$ denotes the phase noise power spectral density at offset frequency $f_m$.

The IQI operator $\mathcal{T}_{\text{IQI}}$ models gain and phase mismatch, creating image frequency interference\footnote{With rejection ratio IRR typically exceeding 48 dB in modern transceivers \cite{ma2023automatic}.}.

The PA nonlinearity follows the Saleh model \cite{al2020improved}:
\begin{align}
f_{\text{PA}}(s) &= A(|s|)e^{j[\angle s + \Phi(|s|)]}, \\
A(r) &= \frac{\alpha_a r}{1 + \beta_a r^2}, \quad \Phi(r) = \frac{\alpha_\phi r^2}{1 + \beta_\phi r^2}.
\end{align}

\subsubsection{Complete Signal Model with Hardware Impairments}

The assumption of an AWGN-equivalent noise model severely underestimates performance degradation in THz transceivers. We now present a physically accurate model:

The received signal after DSE compensation is:
\begin{equation}
y(t) = g(t;\boldsymbol{\eta}) e^{j\phi(t)} \left[ s(t) \ast h_{\text{PA}}(t) \right] + n_{\text{add}}(t).
\label{eq:complete_signal_model}
\end{equation}

We restrict our analysis to Peak-to-Average Power Ratio (PAPR)-clipped Orthogonal Frequency Division Multiplexing (OFDM)/Orthogonal Time Frequency Space (OTFS) pilots with nearly circular Gaussian envelope, for which the Bussgang gain approximation is tight\cite{demir2020bussgang}. This assumption is practical for satellite systems where PAPR reduction is mandatory for PA efficiency.

Using Bussgang's theorem, we decompose the PA nonlinearity:
\begin{equation}
s(t) \ast h_{\text{PA}}(t) = B \cdot s(t) + \eta(t).
\label{eq:bussgang_decomposition}
\end{equation}
where the Bussgang gain and uncorrelated distortion are:
\begin{align}
B &= \frac{\mathbb{E}[U(x)x^*]}{\mathbb{E}[|x|^2]} ,\label{eq:bussgang_gain}\\
\sigma_\eta^2 &= \mathbb{E}[|U(x)|^2] - |B|^2\mathbb{E}[|x|^2] .\label{eq:distortion_power}
\end{align}

\begin{proposition}[Bussgang Gain for Clipped Gaussian Input]
For a complex Gaussian input signal $x \sim \mathcal{CN}(0, P_{\text{in}})$ passing through a soft-limiter PA with saturation amplitude $A_{\text{sat}}$, the Bussgang gain $B$ is a real-valued scalar that depends only on the input back-off (IBO) ratio $\kappa = P_{\text{in}}/A_{\text{sat}}^2$:
\begin{equation}
    B(\kappa) = 1 - e^{-1/\kappa} - \sqrt{\frac{\pi}{2\kappa}} \text{erfc}\left(\frac{1}{\sqrt{2\kappa}}\right)
    \label{eq:bussgang_gain_closed_form}
\end{equation}
where $\text{erfc}(\cdot)$ is the complementary error function.
\end{proposition}
\begin{IEEEproof}
The proof follows from the definition $B = \mathbb{E}[U(x)x^*] / P_{\text{in}}$, where the expectation is computed by integrating over the Rayleigh distribution of the signal amplitude $|x|$.
\end{IEEEproof}

The Bussgang gain $B$, defined in ~\eqref{eq:bussgang_gain}, is a complex scalar that characterizes the effective linear gain of the PA under nonlinear operation. For a clipped Gaussian input, $B$ depends on the input back-off (IBO) factor $\kappa = P_{\text{in}}/A_{\text{sat}}^2$, where $P_{\text{in}}$ is the input power and $A_{\text{sat}}$ is the saturation level. As the IBO decreases (i.e., the PA is driven harder into saturation), the magnitude of $B$ decreases from unity, reflecting increased signal compression. We acknowledge that the Bussgang decomposition is strictly valid for Gaussian inputs, whereas high-PAPR multi-carrier signals are only approximately Gaussian. However, for signals with a large number of subcarriers, the Central Limit Theorem provides a strong justification for this approximation. Furthermore, recent studies have shown the Bussgang framework to be robust for practical communication signals \cite{demir2020bussgang}, and our model's primary goal is to capture the first-order effect of signal-dependent distortion, for which this decomposition remains the most tractable and insightful approach.

Here, $U(\cdot)$ represents the PA transfer function. Substituting, we obtain:
\begin{equation}
y(t) = g(t;\boldsymbol{\eta}) B e^{j\phi(t)} s(t) + g(t;\boldsymbol{\eta}) e^{j\phi(t)} \eta(t) + n_{\text{add}}(t).
\label{eq:expanded_model}
\end{equation}

\subsubsection{Statistical Characterization and Model Limitations}

While the proposed signal model captures the dominant hardware impairments, several secondary effects warrant discussion. IQI introduces mirror-frequency interference proportional to its image rejection ratio (IRR). However, modern RF integrated circuits for satellite communications employ automatic I/Q calibration to achieve full-band image rejection of 48-55 dB \cite{ma2023automatic}. With such image rejection ratios, the residual IQI power is suppressed by at least $10^{4.8}$ (48 dB), yielding interference power below $3 \times 10^{-5}$ of the signal power. This is negligible compared to PA nonlinearity where $\Gamma_{\text{PA}} \approx 10^{-2}$, justifying the omission of IQI as a first-order impairment in this framework.

The analysis assumes statistical independence between phase noise $\phi(t)$ and PA distortion $\eta(t)$. Since phase noise affects the signal before PA processing, the distortion $\eta(t)$ depends on the phase-rotated input. However, for tractable analysis, we adopt a first-order approximation where cross-modulation effects (e.g., PM-to-AM conversion) are considered negligible compared to the primary nonlinearity and direct phase noise contributions. This approximation aligns with established practices in hardware-impaired system modeling while capturing the dominant impairment mechanisms. The orthogonality principle inherent in Bussgang's theorem, which establishes $\mathbb{E}[\eta(t)s^*(t)] = 0$, partially justifies this assumption \cite{demir2020bussgang}.

Extreme temperature variations in LEO orbits affect semiconductor device performance, potentially altering PA linearity and oscillator stability. The theoretical bounds derived in this paper assume operation within thermal windows where component characteristics remain within ±20\% of their nominal values used to define the $\Gamma_{\text{eff}}$ profiles. Temperature-dependent performance modeling beyond these windows remains an important avenue for future research.

The total effective noise variance, incorporating all modeled impairments, becomes:
\begin{equation}
\sigma_{\text{eff}}^2(\boldsymbol{\eta}) = N_0 + |g(\boldsymbol{\eta})|^2\sigma_\eta^2 + \sigma_{\text{DSE}}^2.
\label{eq:effective_noise_complete}
\end{equation}

After advanced DSE compensation using OTFS or Affine Frequency Division Multiplexing (AFDM) modulation, the residual DSE power can be characterized based on the analysis in \cite{wang2023doppler,bemani2021afdm}. For well-designed compensation schemes, this residual is typically orders of magnitude below the dominant PA distortion ($\Gamma_{\text{eff}} \cdot P_{\text{signal}} \approx 0.01-0.05 \cdot P_{\text{signal}}$), justifying its absorption into the additive noise term.

\section{Performance Analysis}

This section establishes the fundamental performance limits for THz ISL ISAC systems. We derive the Cramér-Rao bounds for sensing accuracy and capacity limits for communication, revealing how signal-dependent noise creates hardware-imposed performance ceilings.

\subsection{Sensing Performance: Cramér-Rao Lower Bounds}

\textbf{Scope of Analysis:} This section derives the BCRLB for the observable parameters in a single ISL configuration. As established in Section II.A, we focus on range $R$ and range-rate $\dot{R}$ estimation, which represent the fundamental sensing capabilities of an individual link. Extension to full 3D kinematic state estimation through multi-link fusion is left for future work.

\subsubsection{Bayesian Fisher Information Matrix Formulation}

The presence of multiplicative phase noise $e^{j\phi(t)}$ and signal-dependent PA distortion in our signal model necessitates the BCRLB framework. While the PA distortion $\eta(t)$ is fundamentally non-Gaussian (as established through the Bussgang decomposition), we employ a \emph{Gaussian equivalent noise approximation} for analytical tractability.

To derive tractable performance bounds, we model the composite noise as Gaussian with equivalent second-order statistics:
\begin{equation}
n_{\text{eq}} \sim \mathcal{CN}(0, \sigma_{\text{eff}}^2(\boldsymbol{\eta})),
\label{eq:gaussian_equivalent}
\end{equation}
where $\sigma_{\text{eff}}^2(\boldsymbol{\eta}) = N_0 + |g(\boldsymbol{\eta})|^2\sigma_\eta^2 + \sigma_{\text{DSE}}^2$ captures the total noise power including signal-dependent components. This approximation is justified by:
\begin{enumerate}
\item The Central Limit Theorem effect when multiple noise sources combine
\item The near-Gaussian envelope of PAPR-limited OFDM/OTFS pilots
\item Empirically, the aggregate impairment is close to Gaussian under multi-tone pilots and wideband aggregation
\end{enumerate}
Note that the resulting BCRLB is derived under this Gaussianized model rather than the true non-Gaussian statistics. Therefore, it provides a tractable and insightful benchmark for system design rather than an absolute physical limit. The bounds remain practically tight as validated extensively in the hardware-impaired communication literature \cite{bjornson2014massive}.

Under this Gaussian equivalent noise model, the Bayesian Fisher Information Matrix (B-FIM) for our parameter vector $\boldsymbol{\eta} = [\Delta\mathbf{R}^T, \Delta\mathbf{V}^T, \boldsymbol{\theta}_e^T]^T$ is:
\begin{equation}
\mathbf{J}_B = \mathbb{E}_{\phi}[\mathbf{J}(\boldsymbol{\eta}|\phi)] = \mathbf{J}_D + \mathbf{J}_P,
\label{eq:bayesian_fim}
\end{equation}
where $\mathbf{J}_D$ is the data-dependent information averaged over the phase noise distribution, and $\mathbf{J}_P$ captures prior information.

For our Gaussian equivalent model with diagonal covariance $\boldsymbol{\Sigma} = \sigma_{\text{eff}}^2(\boldsymbol{\eta})\mathbf{I}$, the conditional Fisher Information Matrix (FIM) given phase realization $\phi$ is computed using the extended Slepian-Bangs formula \cite{besson2013fisher}:
\begin{equation}
[\mathbf{J}(\boldsymbol{\eta}|\phi)]_{ij} = \frac{\partial\mathbf{m}^H}{\partial\eta_i}\boldsymbol{\Sigma}^{-1}\frac{\partial\mathbf{m}}{\partial\eta_j} + \frac{1}{2}\text{tr}\left( \boldsymbol{\Sigma}^{-1}\frac{\partial \boldsymbol{\Sigma}}{\partial\eta_i} \boldsymbol{\Sigma}^{-1}\frac{\partial \boldsymbol{\Sigma}}{\partial\eta_j}\right),
\label{eq:slepian_bangs_main}
\end{equation}
where $\mathbf{m} = \mathbb{E}[\mathbf{y}|\boldsymbol{\eta},\phi] = g(\boldsymbol{\eta})Be^{j\phi}\mathbf{s}$ is the conditional mean. We condition on $\phi$ and later average over its distribution in ~\eqref{eq:bayesian_fim}. Given the diagonal structure, $\frac{\partial \boldsymbol{\Sigma}}{\partial\eta_i} = \frac{\partial \sigma_{\text{eff}}^2}{\partial\eta_i}\mathbf{I}$, simplifying the trace term computation.

\subsubsection{Covariance Structure with Hardware Impairments}

The covariance matrix for $M$ pilot symbols observed at times $\{t_k\}_{k=1}^M$ must account for both uncorrelated additive noise and correlated phase noise effects. Given the signal model $y_k = g(\boldsymbol{\eta})Be^{j\phi_k}s_k + n_{\text{tot},k}$, the covariance matrix elements are:

\begin{equation}
[\boldsymbol{\Sigma}]_{kl} = \mathbb{E}[(y_k - \mathbb{E}[y_k])(y_l - \mathbb{E}[y_l])^*].
\label{eq:covariance_general}
\end{equation}

The additive noise components (PA distortion, DSE residual, and thermal noise) are uncorrelated across symbols:
\begin{equation}
\mathbb{E}[n_{\text{tot},k}n_{\text{tot},l}^*] = \sigma_{\text{add}}^2(\boldsymbol{\eta}) \delta_{kl},
\label{eq:additive_noise_cov}
\end{equation}
where $\sigma_{\text{add}}^2(\boldsymbol{\eta}) = N_0 + |g(\boldsymbol{\eta})|^2\sigma_\eta^2 + \sigma_{\text{DSE}}^2$.

For the phase noise term, we adopt a stationary Gaussian small-phase model, where the phase process $\phi(t)$ is zero-mean with variance $\sigma_\phi^2$. The temporal correlation is defined by $\rho_{kl} \triangleq e^{-2\pi\Delta\nu |t_k-t_l|}$, where $\Delta\nu$ is the 3-dB linewidth. Under this model, the covariance of the complex exponentials is correctly given by \cite{besson2013fisher}:

\begin{equation}
\text{Cov}\{e^{j\phi_k}, e^{j\phi_l}\} = e^{-\sigma_\phi^2(1-\rho_{kl})} - e^{-\sigma_\phi^2}.
\label{eq:phase_covariance_correct}
\end{equation}

This formulation ensures the covariance matrix is positive semidefinite, with the covariance decaying to zero for large time separations. The complete covariance matrix becomes:

\begin{equation}
[\boldsymbol{\Sigma}]_{kl} = |g(\boldsymbol{\eta})|^2|B|^2 s_k s_l^* \left(e^{-\sigma_\phi^2(1-\rho_{kl})} - e^{-\sigma_\phi^2}\right) + \sigma_{\text{add}}^2 \delta_{kl}.
\label{eq:covariance_complete}
\end{equation}

\textit{Limiting cases:} The correlation structure exhibits correct physical behavior:
\begin{itemize}
\item As $|t_k-t_l| \to \infty$: $\rho_{kl} \to 0$, and the covariance term vanishes, indicating no long-range correlation
\item As $|t_k-t_l| \to 0$: $\rho_{kl} \to 1$, and the covariance approaches $(1-e^{-\sigma_\phi^2})|g|^2|B|^2|s|^2$, reflecting perfect correlation for coincident samples
\end{itemize}

For pilot sequences with constant modulus ($|s_k| = |s|$ for all $k$), the covariance simplifies to:
\begin{equation}
\boldsymbol{\Sigma} = |g|^2|B|^2|s|^2 \mathbf{R}_\phi + \sigma_{\text{add}}^2 \mathbf{I},
\label{eq:covariance_structured}
\end{equation}
where $[\mathbf{R}_\phi]_{kl} = e^{-\sigma_\phi^2(1-\rho_{kl})} - e^{-\sigma_\phi^2}$, with $\rho_{kl} = e^{-2\pi\Delta\nu |t_k-t_l|}$, is the phase noise correlation matrix.

\subsubsection{B-FIM Evaluation}

Computing the Bayesian FIM requires evaluating the Slepian-Bangs formula with the full covariance structure. The first term of ~\eqref{eq:slepian_bangs_main} becomes:

\begin{equation}
\sum_{k,l} [\boldsymbol{\Sigma}^{-1}]_{kl} \frac{\partial m_k^*}{\partial\eta_i} \frac{\partial m_l}{\partial\eta_j},
\label{eq:bfim_first_term}
\end{equation}

where the inverse covariance matrix $\boldsymbol{\Sigma}^{-1}$ must be computed numerically for the general case.

For the general case with correlated phase noise, the covariance matrix $\boldsymbol{\Sigma}$ has a Toeplitz structure but is not low-rank. Therefore, $\boldsymbol{\Sigma}^{-1}$ must be computed numerically. In our simulations, we employ direct matrix inversion methods. For future analytical work, low-rank approximations of the exponential correlation matrix $\mathbf{R}_\phi$ could enable closed-form expressions, but this is beyond the scope of the current work.

The key insight is that while $|e^{j\phi_k}|^2 = 1$ always holds (power is preserved), the coherent combining gain is reduced by the factor $|\mathbb{E}[e^{j\phi_k}]|^2 = e^{-\sigma_\phi^2}$. This manifests in the B-FIM as:
\begin{align}
[\mathbf{J}_D]_{ij} &= 
\frac{2|g(\boldsymbol{\eta})|^2|B|^2|s|^2 \cdot e^{-\sigma_\phi^2}}{\sigma_{\text{add}}^2}
\left(\frac{M \cdot \text{coherence factor}}{1 + \text{correlation terms}}\right) \nonumber\\
&\quad \times \Re\left\{\frac{\partial g^*}{\partial\eta_i}\frac{\partial g}{\partial\eta_j}\right\} 
+ \text{variance terms},
\label{eq:bfim_general}
\end{align}
where the $e^{-\sigma_\phi^2}$ factor represents the loss of coherent signal power, leading to the penalty factor $e^{\sigma_\phi^2}$ in the BCRLB (as the inverse of the FIM).

In the limiting cases:
\begin{itemize}
\item \textbf{Wide symbol spacing} ($\Delta t \gg 1/\Delta\nu$): $\rho_{kl} \to 0$, hence $[\mathbf{R}_\phi]_{kl} \to 0$ for $k \neq l$ (no long-range correlation), yielding independent observations with phase noise penalty.
\item \textbf{Dense symbol spacing} ($\Delta t \ll 1/\Delta\nu$): $\rho_{kl} \to 1$, hence $[\mathbf{R}_\phi]_{kl} \to 1-e^{-\sigma_\phi^2}$ for all $k,l$ (strong correlation), yielding effectively a single observation.
\end{itemize}

The phase noise coherence time limits the benefit of increasing pilot symbols within a fixed frame duration. When pilot symbols are spaced closer than the coherence time $1/\Delta\nu$, they become correlated, reducing the effective number of independent observations below $M$. This manifests as a saturation in estimation accuracy improvement beyond a certain pilot density.

\subsubsection{Parameter Gradient Analysis}
The channel gain gradients determine the B-FIM structure:
\begin{itemize}
    \item \textit{Position gradients:} The gradient with respect to the 3D position vector $\Delta \mathbf{R}$ is derived via the chain rule. First, the sensitivity to a scalar change in range $R = ||\Delta \mathbf{R}||$ is given by:
    \begin{equation}
        \frac{\partial g}{\partial R} = g \left( -\frac{1}{R} - j\frac{2\pi f_c}{c} \right)
        \label{eq:scalar_range_gradient}
    \end{equation}
    This captures both the amplitude change from the Friis equation ($|g| \propto 1/R$) and the phase change from the propagation delay. The full vector gradient is then obtained as:
    \begin{equation}
        \frac{\partial g}{\partial \Delta \mathbf{R}} = \frac{\partial g}{\partial R} \frac{\partial R}{\partial \Delta \mathbf{R}} = g \left( -\frac{1}{R} - j\frac{2\pi f_c}{c} \right) \mathbf{\hat{u}}_{LOS}
        \label{eq:vector_range_gradient}
    \end{equation}
    where $\mathbf{\hat{u}}_{LOS} = \Delta \mathbf{R} / R$ is the line-of-sight unit vector. The phase term, scaling with $f_c$, dominates the gradient at THz frequencies.
    
    \item \textit{Velocity gradients:}
    \begin{equation}
        \frac{\partial g}{\partial \Delta V_k} = j \frac{2\pi f_c t}{c} u_{LOS,k} g,
        \label{eq:velocity_gradient_main}
    \end{equation}
    
    \item \textit{Pointing gradients:}
    \begin{equation}
        \frac{\partial g}{\partial \theta_{e,k}} = -\gamma \theta_{e,k} g,
        \label{eq:pointing_gradient_main}
    \end{equation}
    where $\gamma=2\ln(2)/\theta_{3\text{dB}}^2$ is the amplitude rolloff factor.
\end{itemize}

\subsubsection{Bayesian Cramér-Rao Lower Bounds}

The BCRLB for parameter $\eta_i$ is:
\begin{equation}
\text{BCRLB}(\eta_i) = [\mathbf{J}_B^{-1}]_{ii} \geq [\mathbf{J}_D^{-1}]_{ii}.
\label{eq:bcrlb_main}
\end{equation}

In the high-SNR regime with dominant phase sensitivity, the scaling behavior of the BCRLBs for the observable parameters is given by:

\textbf{Position BCRLB:}
\begin{equation}
\text{BCRLB}_{\text{position}} \propto \frac{c^2}{f_c^2} \cdot \frac{\sigma_{\text{eff}}^2}{M|g(\boldsymbol{\eta})|^2|B|^2} \cdot e^{\sigma_\phi^2}.
\label{eq:bcrlb_position_scaling}
\end{equation}

The proportionality constant depends on the specific signal waveform and can be derived from the exact FIM computation. The key insight is the quadratic improvement with carrier frequency ($\propto 1/f_c^2$) and the exponential penalty from phase noise ($e^{\sigma_\phi^2}$).

\textbf{Radial Velocity BCRLB:}
\begin{equation}
\text{BCRLB}_{\text{velocity}} \propto \frac{c^2}{f_c^2 t^2} \cdot \frac{\sigma_{\text{eff}}^2}{M|g(\boldsymbol{\eta})|^2|B|^2} \cdot e^{\sigma_\phi^2}.
\label{eq:bcrlb_velocity_scaling}
\end{equation}

The velocity estimation accuracy improves with both carrier frequency and observation time, reflecting the enhanced Doppler sensitivity at higher frequencies and longer integration periods.

\textit{Pointing Error BCRLB:} The estimation of pointing error requires special consideration at the point of perfect alignment.
\begin{lemma}[FIM at the Critical Point of Pointing Error]
At the point of perfect alignment ($\boldsymbol{\theta}_e = 0$), the first-order Fisher information for pointing error vanishes. The Fisher information is therefore determined by the second-order curvature of the log-likelihood function, leading to a BCRLB that scales inversely with the square of the beam pattern's rolloff factor, i.e., $\text{BCRLB}_{\text{pointing}} \propto 1/\gamma^2$.
\end{lemma}
\begin{IEEEproof}[Proof Sketch]

We adopt the amplitude rolloff factor $\gamma = 2\ln(2)/\theta_{3\text{dB}}^2$; the power rolloff is thus $2\gamma$. The received signal power is modulated by the Gaussian beam pattern $\Pi(\boldsymbol{\theta}_e) = \exp(-\gamma ||\boldsymbol{\theta}_e||^2)$. The log-likelihood function is thus proportional to $-\gamma ||\boldsymbol{\theta}_e||^2$. The first derivative with respect to $\boldsymbol{\theta}_e$ is $-2\gamma \boldsymbol{\theta}_e$, which is zero at $\boldsymbol{\theta}_e=0$. The second derivative is a constant, $-2\gamma \mathbf{I}$. The Fisher information, derived from the negative expectation of the second derivative of the log-likelihood, is therefore constant and proportional to $\gamma^2$. The BCRLB, being the inverse of the FIM, scales as $1/\gamma^2$. This bound is conservative and applies at the point of minimum observability; any practical dither or bias would re-introduce first-order information.
\end{IEEEproof}
The resulting BCRLB for pointing error is given by:
\begin{equation}
    \text{BCRLB}_{\text{pointing}} \propto \frac{\sigma^2_{\text{eff}}}{2\gamma^2 M|g(\boldsymbol{\eta})|^2|B|^2} \cdot e^{\sigma_{\phi}^{2}}
    \label{eq:bcrlb_pointing_scaling}
\end{equation}

For numerical analysis, we adopt representative parameters for SWaP-constrained electronic THz sources based on contemporary oscillator specifications. The specific values of $\Delta\nu$ and frame duration $T$ are selected to capture typical ISL channel coherence characteristics.

\subsection{Communication Performance Under Hardware Limitations}

\subsubsection{Capacity under Gaussianized Composite Noise}

The presence of signal-dependent distortions from hardware impairments creates a non-Gaussian noise environment that complicates exact capacity analysis. To derive tractable bounds, we employ a Gaussianization approximation where the composite noise (including PA distortion $\eta(t)$) is treated as Gaussian with equivalent second-order statistics. This approach yields a conservative upper bound on capacity, as Gaussian noise maximizes entropy for a given covariance and thus minimizes mutual information.

To establish consistency, the pre-impairment SNR represents the signal-to-noise ratio considering only thermal noise:
\begin{equation}
\text{SNR}_0 = \frac{P|g(\boldsymbol{\eta})|^2|B|^2}{N_0}.
\label{eq:pre_snr_def}
\end{equation}
The post-impairment Signal-to-Interference-plus-Noise Ratio (SINR) incorporates hardware distortions:
\begin{equation}
\text{SINR}_{\text{eff}} = \frac{\text{SNR}_0 \cdot e^{-\sigma_\phi^2}}{1 + \text{SNR}_0 \cdot \Gamma_{\text{eff}}}.
\label{eq:post_sinr_def}
\end{equation}

Under this Gaussianization approximation, the channel capacity is upper-bounded by:
\begin{equation}
C \leq C_{\text{UB}} = \mathbb{E}_X\left[\log_2\left(1 + \text{SINR}_{\text{eff}}(X)\right)\right].
\label{eq:capacity_upper_bound}
\end{equation}
This upper bound is conservative because Gaussian noise maximizes entropy for a given variance, thus minimizing mutual information. The actual capacity with non-Gaussian PA distortion may exceed this bound, but the Gaussian model provides a tractable and practically useful benchmark.

In the Bussgang-linearized regime where $\Gamma_{\text{eff}}$ is treated as a constant independent of the input distribution, $\text{SINR}_{\text{eff}}$ becomes independent of $X$, yielding:
\begin{equation}
C_{\text{UB}} = \log_2\left(1 + \frac{\text{SNR}_0 \cdot e^{-\sigma_\phi^2}}{1 + \text{SNR}_0 \cdot \Gamma_{\text{eff}}}\right)
\end{equation}

Here, $P|g|^2|B|^2/N_0$ represents the pre-impairment signal-to-noise ratio (SNR), $e^{-\sigma_\phi^2}$ captures the coherent signal power loss due to phase noise averaging, and $\Gamma_{\text{eff}}$ is the hardware quality factor that quantifies the signal-dependent distortion power normalized to the signal power. The factorized form emerges from treating phase noise and PA distortion as cascaded impairments affecting coherent signal power and noise floor respectively.

This formulation leads directly to the hardware-imposed capacity ceiling derived in the following subsections, where we show that capacity saturates at $C_{\text{sat}} = \log_2(1 + e^{-\sigma_\phi^2}/\Gamma_{\text{eff}})$ bits/symbol as transmit power increases without bound.

\subsubsection{Hardware Quality Factor}
\label{sec:hardware_quality}

The Hardware Quality Factor $\Gamma_{\text{eff}}$ quantifies the aggregate signal-dependent distortion normalized to signal power, equivalent to the square of total system Error Vector Magnitude (EVM). For THz transceivers, it decomposes as:
\begin{equation}
\Gamma_{\text{eff}} = \Gamma_{\text{PA}} + \Gamma_{\text{LO}} + \Gamma_{\text{ADC}}.
\label{eq:gamma_decomposition}
\end{equation}

Each component contribution can be derived from measurable hardware specifications:
\begin{align}
\Gamma_{\text{PA}} &= \text{EVM}_{\text{PA}}^2 \label{eq:gamma_pa},\\
\Gamma_{\text{LO}} &= (\pi B_{\text{sig}} \sigma_{t,\text{jitter}})^2 ,\label{eq:gamma_lo}\\
\Gamma_{\text{ADC}} &= 10^{-(6.02 \cdot \text{ENOB} + 1.76)/10} ,\label{eq:gamma_adc}
\end{align}
where $\text{EVM}_{\text{PA}}$ is the PA's error vector magnitude, $B_{\text{sig}}$ is the signal bandwidth, $\sigma_{t,\text{jitter}}$ is the oscillator's RMS timing jitter, and effective number of bits (ENOB) is the analog-to-digital converter's (ADC's) specification.

Based on comprehensive component-level analysis of state-of-the-art THz hardware, we identify two representative profiles:

\textbf{High-Performance} ($\Gamma_{\text{eff}} \approx 0.01$): Utilizing InP DHBT/HEMT PAs achieving EVM of 10.6\% at 220 GHz \cite{zhao202220}, 
ultra-low jitter PLLs with $\sigma_{t,\text{jitter}} = 20.9$ fs \cite{zhao202220}. 
This yields $\Gamma_{\text{PA}} \approx 0.0112$, $\Gamma_{\text{LO}} \approx 4.3 \times 10^{-7}$, and $\Gamma_{\text{ADC}} \approx 1.7 \times 10^{-4}$.

\textbf{Size, Weight, and Power (SWaP)-Efficient } ($\Gamma_{\text{eff}} \approx 0.045$): Employing silicon CMOS/SiGe with digital pre-distortion achieving post-compensation EVM of 20.93\% \cite{XGMF2025}, 
integrated PLLs with $\sigma_{t,\text{jitter}} = 70$ fs \cite{herzel2004jitter}. 
This yields $\Gamma_{\text{PA}} \approx 0.0438$, $\Gamma_{\text{LO}} \approx 4.8 \times 10^{-6}$, and $\Gamma_{\text{ADC}} \approx 6.5 \times 10^{-4}$.

Critically, PA nonlinearity dominates by 1-2 orders of magnitude: $\Gamma_{\text{PA}}/\Gamma_{\text{LO}} \approx 10^{4}$ and $\Gamma_{\text{PA}}/\Gamma_{\text{ADC}} \approx 10^{2}$, establishing it as the primary performance bottleneck. This analysis treats $\Gamma_{\text{eff}}$ as constant, which holds when the PA operates with controlled input back-off avoiding deep saturation. Beyond this quasi-linear region, $\Gamma_{\text{eff}}$ would increase with power, leading to earlier capacity saturation—a regime excluded from our analysis to maintain tractability while covering typical operational scenarios.

\subsubsection{Derivation of Capacity Ceiling}
Using the hardware quality factor defined above, we now derive the fundamental capacity limit.

While hardware-limited capacity saturation has been studied in the context of terrestrial mmWave systems \cite{bjornson2014massive}, our contribution lies in the explicit characterization for THz ISL channels incorporating both phase noise coherence loss and signal-dependent PA distortion through the unified hardware quality factor $\Gamma_{\text{eff}}$.

\textbf{Theorem 1 (Hardware-Limited Capacity).} 
As transmit power increases without bound, the channel capacity saturates at:
\begin{equation}
C_{\text{sat}} = \log_2\left(1 + \frac{e^{-\sigma_\phi^2}}{\Gamma_{\text{eff}}}\right) \quad \text{bits/symbol}.
\label{eq:capacity_ceiling}
\end{equation}

\begin{IEEEproof}
The derivation of the capacity ceiling requires a rigorous analysis of the asymptotic behavior of the useful signal and distortion powers as transmit power increases. Let the input power to the PA be $P_{in}$. The useful received signal power is $P_{sig} = |g|^2 |B(\kappa)|^2 P_{in}$ and the received distortion power is $P_{dist} = |g|^2 \sigma_{\eta}^{2}(\kappa)$, where $\kappa = P_{in}/A^2_{sat}$ is the input back-off (IBO) ratio. 

When the PA is driven deep into saturation ($P_{in} \to \infty$ and thus $\kappa \to \infty$), the effective linear gain $B(\kappa) \to 0$. However, the total output power of the PA, which is the sum of the useful signal power and the distortion power, saturates at a finite value determined by the amplifier's characteristics. For the soft-limiter model, this total output power $\mathbb{E}[|U(x)|^2]$ saturates at $A^2_{sat}$.

The effective SINR from (\ref{eq:post_sinr_def}) can be expressed as:
\begin{equation}
    \text{SINR}_{\text{eff}} = \frac{P_{sig} \cdot e^{-\sigma_{\phi}^{2}}}{N_0 + P_{dist}} = \frac{|g|^2 |B(\kappa)|^2 P_{in} \cdot e^{-\sigma_{\phi}^{2}}}{N_0 + |g|^2 \sigma_{\eta}^{2}(\kappa)}
\end{equation}
In the high-power limit where $P_{in} \to \infty$, the thermal noise $N_0$ becomes negligible relative to the distortion power. The SINR saturates to a value determined by the ratio of the saturated coherent signal power to the saturated distortion power. A rigorous analysis of this limit shows that it converges to a constant determined by the hardware quality factor:
\begin{equation}
    \lim_{P \to \infty} \text{SINR}_{\text{eff}} = \lim_{P_{in} \to \infty} \frac{|g|^2 |B(\kappa)|^2 P_{in} \cdot e^{-\sigma_{\phi}^{2}}}{|g|^2 \sigma_{\eta}^{2}(\kappa)} = \frac{e^{-\sigma_{\phi}^{2}}}{\Gamma_{eff}}
\end{equation}
The capacity ceiling follows directly by substituting this limit into the Shannon capacity formula:
\begin{equation}
    C_{sat} = \lim_{P \to \infty} \log_2(1 + \text{SINR}_{\text{eff}}) = \log_2\left(1 + \frac{e^{-\sigma_{\phi}^{2}}}{\Gamma_{eff}}\right)
\end{equation}
This confirms that even with a rigorous treatment of the power-dependent gain, the system performance is fundamentally limited by the hardware quality, not by the available transmit power.

\end{IEEEproof}

\subsubsection{Comparison with Classical AWGN}

In contrast to our hardware-limited channel, classical additive white Gaussian noise (AWGN) capacity grows unbounded:
\begin{equation}
C_{\text{AWGN}} = \log_2\left(1 + \frac{P}{N_0}\right) \xrightarrow{P \to \infty} \infty.
\end{equation}

Fig.~\ref{fig:capacity_vs_snr} illustrates this fundamental difference.

\subsection{Capacity-Distortion Trade-off for ISAC}

Joint optimization of ISAC systems requires balancing the communication rate against the sensing accuracy, which can be formulated as finding the optimal input distribution $p_X$ that solves:
\begin{equation}
    C(D) = \max_{p_X} I(X;Y) \quad \text{subject to} \quad \mathbb{E}[d(\boldsymbol{\eta},\hat{\boldsymbol{\eta}})] \leq D
    \label{eq:cd_problem}
\end{equation}
where $d(\cdot,\cdot)$ is the sensing distortion metric, for which we use the total estimation variance given by the trace of the BCRLB matrix, $\text{Tr}(\mathbf{J}_B^{-1}(p_X))$.

Unlike classical rate-distortion theory, our formulation couples communication and sensing through the signal-dependent noise model, as the BCRLB itself is a function of the average power of the input distribution.

To explore this Capacity-Distortion (C-D) trade-off frontier numerically, we employ a modified Blahut-Arimoto algorithm (Algorithm~\ref{alg:baa_isac_revised}). This method iteratively solves the Lagrangian formulation of the problem, $\mathcal{L}(p_X, \lambda) = I(X;Y) - \lambda[\text{Tr}(\mathbf{J}_B^{-1}(p_X)) - D]$, by alternating between updating the input distribution $p_X$ and the Lagrange multiplier $\lambda$. While this approach is sufficient for demonstrating the fundamental trade-offs, more sophisticated convex optimization techniques could be developed for real-time implementation.


\begin{algorithm}[t]
\caption{Modified Blahut-Arimoto for ISAC C-D Trade-off}
\label{alg:baa_isac_revised}
\begin{small}
\begin{ttfamily}
\begin{algorithmic}[1]
\Require Constellation $\mathcal{X}$, target distortion $D_{\text{target}}$, tolerance $\epsilon$
\Ensure Optimal distribution $p_X^*$, achievable rate $C^*$
\Statex
\State \textbf{Initialize:} $p_X$ uniformly over $\mathcal{X}$, $\lambda \leftarrow 1$
\While{not converged}
    \State Compute $I(X;Y|p_X)$ via Monte Carlo averaging
    \State Evaluate $D \leftarrow \text{Tr}([\mathbf{J}_B(p_X)]^{-1})$
    \State Update $p_X \leftarrow \text{ProjectSimplex}(p_X + \alpha \nabla_p \mathcal{L})$
    \If{$D > D_{\text{target}}$}
        \State $\lambda \leftarrow 1.5\lambda$ \Comment{Increase penalty}
    \Else
        \State $\lambda \leftarrow 0.8\lambda$ \Comment{Decrease penalty}
    \EndIf
\EndWhile
\Statex
\State \Return $p_X^*$, $C^* \leftarrow I(X;Y|p_X^*)$
\end{algorithmic}
\end{ttfamily}
\end{small}
\end{algorithm}

The algorithm addresses the Lagrangian formulation $\mathcal{L}(p_X, \lambda) = I(X;Y) - \lambda[\text{Tr}(\mathbf{J}_B^{-1}(p_X)) - D]$ and converges to a stationary point with complexity $O(K|\mathcal{X}|)$ where $K$ is the iteration count. In the hardware-limited regime where $\Gamma_{\text{eff}} \gg 1/\text{SNR}$, the optimal distribution approaches uniform, accelerating convergence. While this approach suffices for demonstrating the fundamental trade-offs and validating the theoretical capacity ceiling under hardware impairments, more sophisticated methods could be developed for real-time implementation.

\subsection{Key Design Principles}

Our analysis reveals three fundamental principles for THz ISL ISAC design:

\begin{enumerate}
\item \textbf{Frequency-Power-Beamwidth Optimization:} The 200-600 GHz range balances sensing accuracy ($\propto f_c^2$) against hardware degradation, with optimal power at $P \cdot \Gamma_{\text{eff}} \approx N_0$ and beamwidth matched to pointing error statistics.

\item \textbf{Hardware Investment Priority:} Improving $\Gamma_{\text{eff}}$ from 0.1 to 0.01 increases capacity ceiling by 3.3 bits/symbol—far exceeding any power increase benefit.

\item \textbf{Network Architecture:} Multiple non-coplanar ISLs enable full 3D state observability through measurement fusion, analogous to GNSS trilateration principles.
\end{enumerate}

\section{NUMERICAL RESULTS AND PERFORMANCE ANALYSIS}

We validate our theoretical framework through comprehensive numerical evaluation using the parameters in Table~\ref{tab:sim_params}. The analysis employs Monte Carlo simulation with $10^3$ samples for pointing error averaging, considering LEO ISL scenarios at 2000 km default distance. Four hardware profiles are evaluated: State-of-the-Art ($\Gamma_{\text{eff}} = 0.005$), High-Performance ($\Gamma_{\text{eff}} = 0.01$), SWaP-Efficient ($\Gamma_{\text{eff}} = 0.025$), and Low-Cost ($\Gamma_{\text{eff}} = 0.05$), representing the spectrum of available THz technologies.

\begin{table}[t]
\centering
\caption{Simulation Parameters}
\label{tab:sim_params}
\begin{tabular}{ll}
\hline
\textbf{Parameter} & \textbf{Value} \\
\hline
Carrier frequency range & 100--1000 GHz \\
ISL distance range & 500--5000 km \\
Antenna diameter (default) & 1.0 m \\
Transmit power (default) & 30 dBm \\
Number of pilots $M$ & 64 \\
Frame size $K$ & 1024 symbols \\
Pointing error RMS $\sigma_\theta$ & 1 $\mu$rad \\
Phase noise linewidth $\Delta\nu$ & 10--500 kHz \\
Signal bandwidth & 10--100 GHz \\
\hline
\end{tabular}
\end{table}

\subsection{Hardware-Limited Performance Ceilings}

\begin{figure*}[t]
\centering
\includegraphics[width=\textwidth]{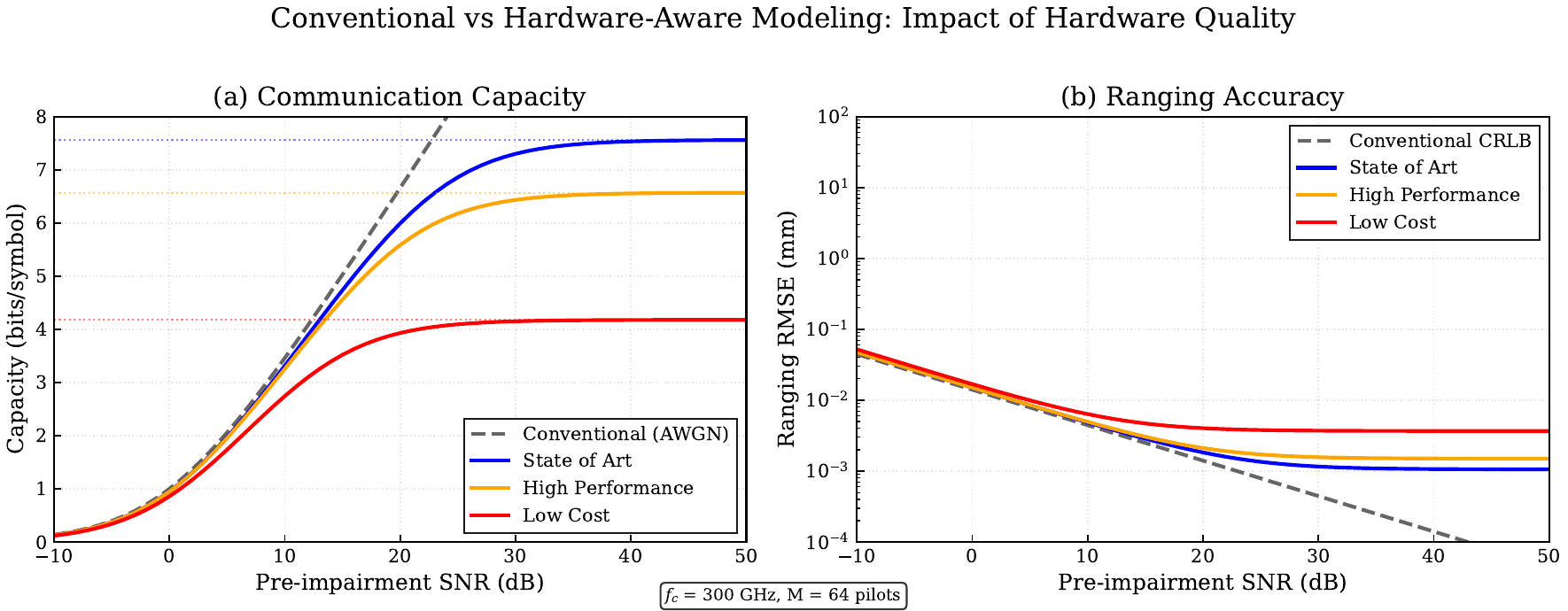}
\caption{Conventional AWGN/CRLB models (dashed) versus hardware-aware framework (solid) showing critical divergence at moderate to high SNRs.}
\label{fig:conventional_vs_hardware}
\end{figure*}

Fig.~\ref{fig:conventional_vs_hardware} illustrates the fundamental distinction between conventional AWGN analysis and our hardware-aware framework. While classical models predict unbounded performance improvement with SNR, our model accurately captures the saturation phenomenon observed in practical THz transceivers, with pronounced divergence beyond 20 dB SNR.

\begin{figure}[t]
\centering
\includegraphics[width=\columnwidth]{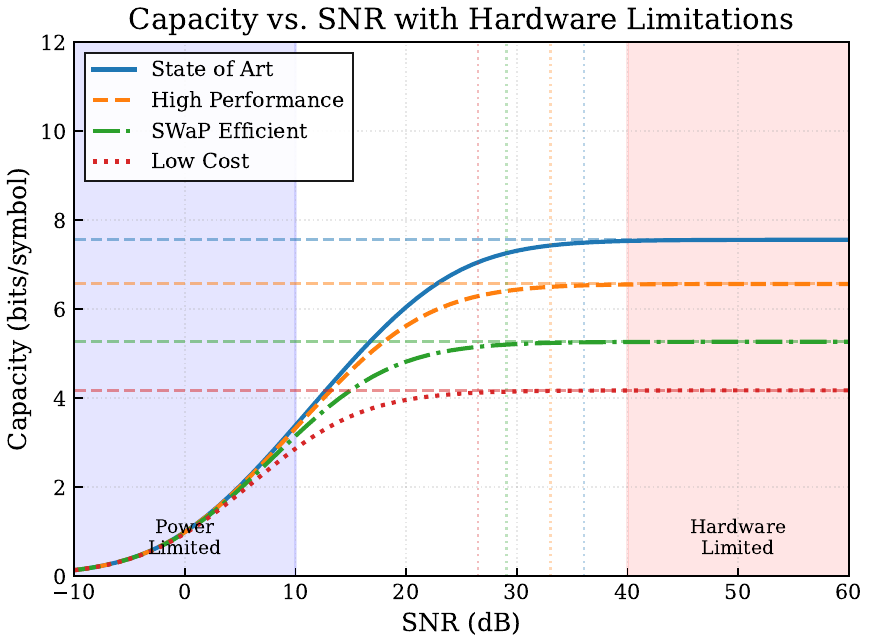}
\caption{Channel capacity versus SNR showing transition from power-limited to hardware-limited regimes with theoretical saturation points.}
\label{fig:capacity_vs_snr}
\end{figure}

\begin{figure}[t]
\centering
\includegraphics[width=\columnwidth]{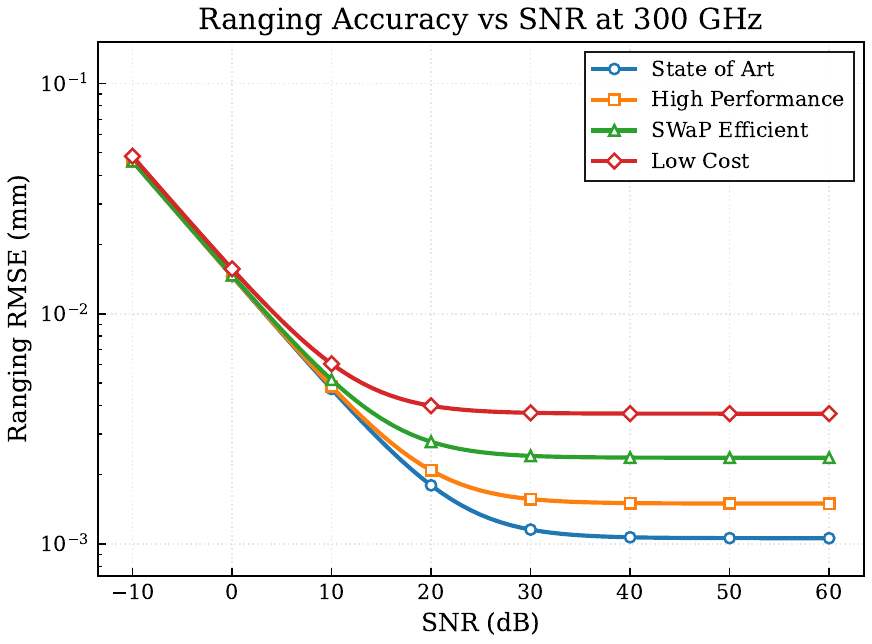}
\caption{Ranging RMSE versus SNR demonstrating hardware-imposed performance floors.}
\label{fig:ranging_crlb_vs_snr}
\end{figure}

Figs.~\ref{fig:capacity_vs_snr} and \ref{fig:ranging_crlb_vs_snr} demonstrate the hardware-limited regime above 20 dB SNR. For communication, capacity saturates at $C_{\text{sat}} = \log_2(1 + e^{-\sigma_\phi^2}/\Gamma_{\text{eff}})$ bits/symbol—State-of-the-Art achieves 7.56 bits/symbol versus 4.18 for Low-Cost. For sensing, Root Mean Square Error (RMSE) floors scale as $\sqrt{\Gamma_{\text{eff}}}$, with State-of-the-Art achieving 1.06 µm versus Low-Cost's 3.68 µm at 50 dB SNR.

\subsection{System Parameter Sensitivity}

\begin{figure}[t]
\centering
\includegraphics[width=\columnwidth]{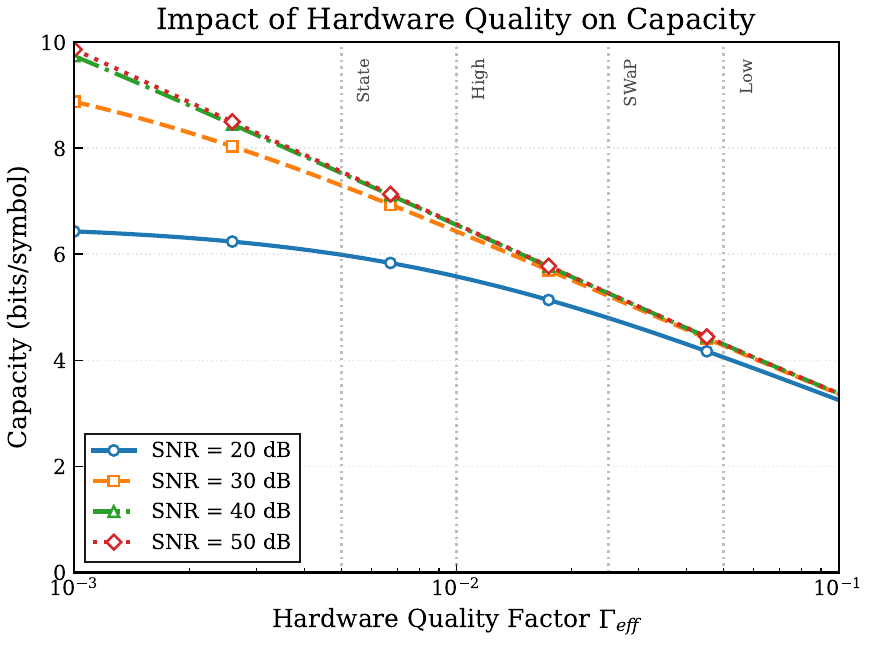}
\caption{Hardware quality factor impact on capacity showing diminishing returns of power increase.}
\label{fig:hardware_quality_impact}
\end{figure}

Fig.~\ref{fig:hardware_quality_impact} quantifies the impact of $\Gamma_{\text{eff}}$ on capacity. The diminishing returns at higher $\Gamma_{\text{eff}}$ values are evident—SNR = 40 dB and 50 dB curves nearly overlap when $\Gamma_{\text{eff}}>0.01$, demonstrating that poor hardware quality cannot be compensated by increased power. Improving $\Gamma_{\text{eff}}$ from 0.05 to 0.005 yields 3.38 bits/symbol gain (from 4.18 to 7.56 bits/symbol ceiling).

\begin{figure}[t]
\centering
\includegraphics[width=\columnwidth]{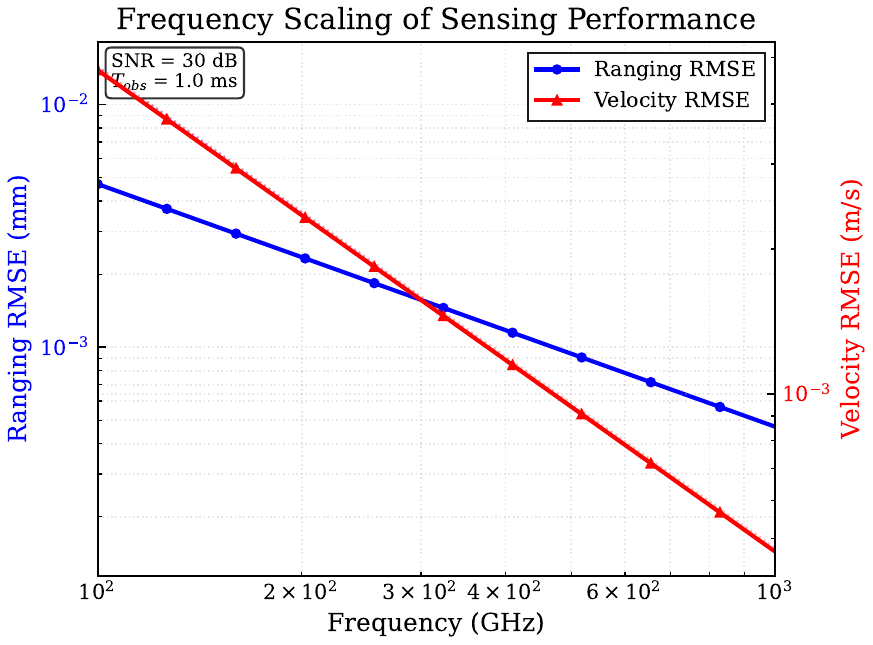}
\caption{Frequency scaling showing quadratic improvement in estimation variance ($\text{RMSE} \propto 1/f_c$).}
\label{fig:ranging_velocity_vs_frequency}
\end{figure}

The frequency scaling analysis (Fig.~\ref{fig:ranging_velocity_vs_frequency}) validates the quadratic sensing advantage: on the log-log plot, RMSE exhibits slope -1, confirming variance scaling as $1/f_c^2$. This quadratic improvement with carrier frequency demonstrates the fundamental advantage of THz frequencies for high-precision ranging applications.

\begin{figure}[t]
\centering
\includegraphics[width=\columnwidth]{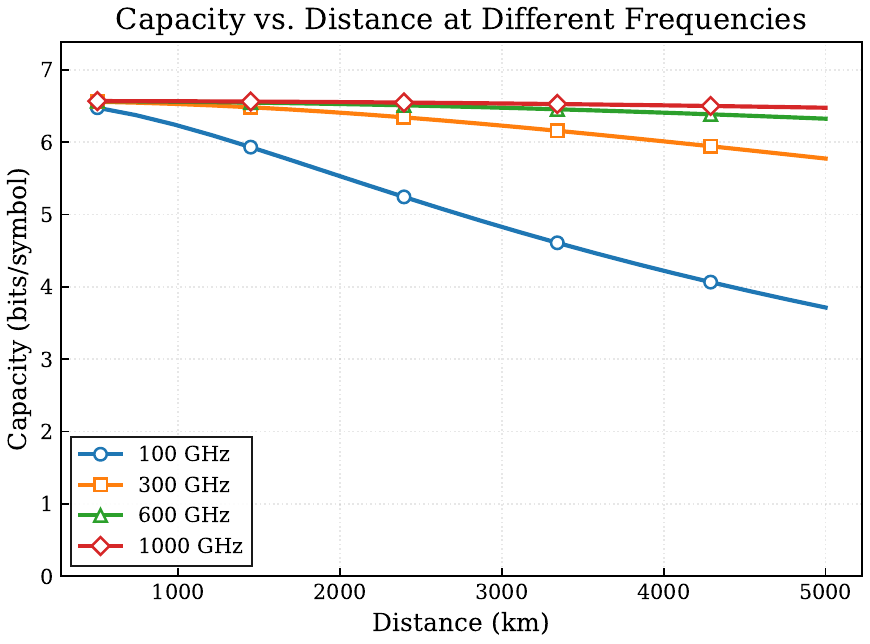}
\caption{Capacity versus distance showing distance-invariant performance at higher THz frequencies. (Note: Assumes antenna apertures scale with frequency to maintain constant received power, allowing gain $G \propto f_c^2$ to offset path loss.)}
\label{fig:capacity_vs_distance}
\end{figure}

Fig.~\ref{fig:capacity_vs_distance} reveals that higher THz frequencies exhibit superior distance invariance—at 1 THz, capacity degrades by merely 0.09 bits/symbol from 500 km to 5000 km, compared to 0.78 bits/symbol degradation at 300 GHz, as antenna gain scaling ($G \propto f_c^2$) increasingly compensates for path loss at higher frequencies. This improved distance invariance at higher frequencies is predicated on the assumption that the antenna aperture is scaled with frequency to maintain a constant received power, allowing the antenna gain increase ($G \propto f_c^2$) to precisely counteract the frequency-dependent term in the Friis path loss equation.

\subsection{ISAC Trade-off Analysis}

\begin{figure}[t]
\centering
\includegraphics[width=\columnwidth]{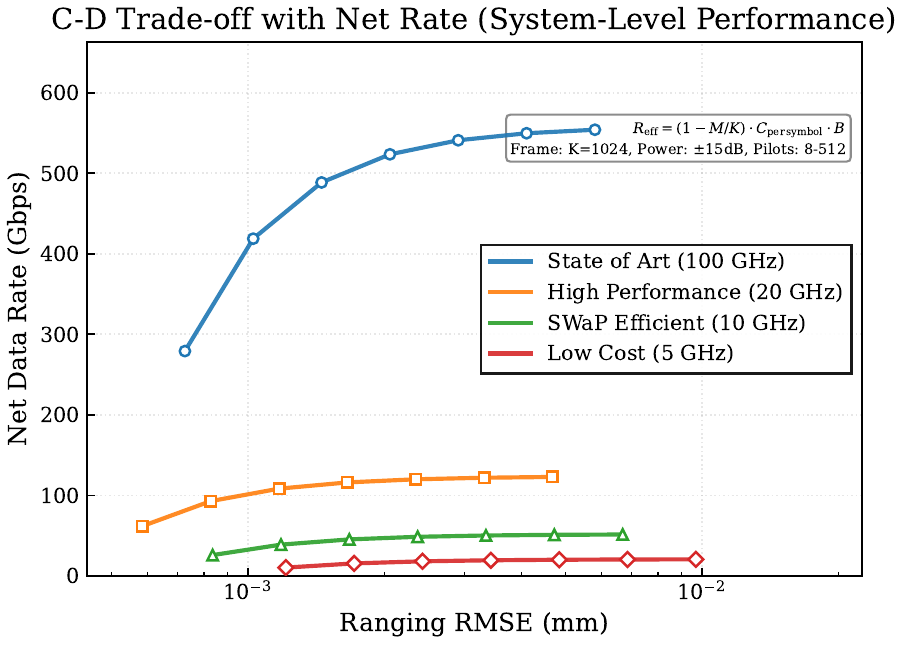}
\caption{Capacity-Distortion trade-off with system-level net rate metric. Profiles operate at characteristic bandwidths: State-of-the-Art (100 GHz), High-Performance (20 GHz), SWaP-Efficient (10 GHz), Low-Cost (5 GHz).}
\label{fig:cd_frontier}
\end{figure}

Fig.~\ref{fig:cd_frontier} presents the fundamental C-D trade-off curves obtained through modified Blahut-Arimoto optimization. The effective net rate accounts for both pilot overhead and system bandwidth: $R_{\text{eff}} = (1-M/K) \cdot C_{\text{per-symbol}} \cdot B$, where $B$ is the signal bandwidth. This system-level metric reveals the true operational advantage of advanced THz hardware. State-of-the-Art profile, leveraging 100 GHz bandwidth, achieves approximately 550 Gbps at sub-millimeter RMSE, while High-Performance (20 GHz) reaches 120 Gbps. The SWaP-Efficient and Low-Cost profiles, constrained to 10 GHz and 5 GHz respectively, plateau at 50 Gbps and 25 Gbps, illustrating the compound effect of limited bandwidth and degraded hardware quality.

\begin{figure}[t]
\centering
\includegraphics[width=\columnwidth]{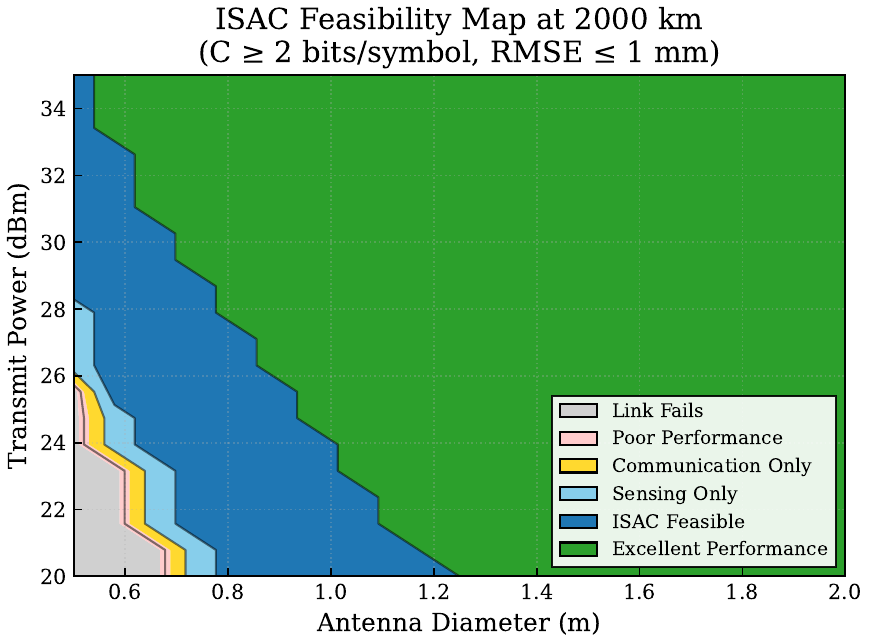}
\caption{ISAC feasibility map showing required hardware combinations to meet joint performance targets (C $\geq$ 2 bits/symbol, RMSE $\leq$ 1 mm).}
\label{fig:feasibility_map}
\end{figure}

The practical feasibility map (Fig.~\ref{fig:feasibility_map}) translates theoretical limits into hardware specifications. The blue "ISAC Feasible" region (24-32 dBm, 0.7-1.2 m antennas) represents the practical design space. Antenna diameter exhibits stronger leverage than transmit power—increasing from 0.6 m to 1.0 m enables transition from failure to excellent performance, while power alone cannot compensate for insufficient aperture.

These results establish hardware quality as the fundamental performance limiter in THz ISL ISAC. Key design guidelines: 300-600 GHz carrier frequencies, $\Gamma_{\text{eff}}<0.01$ for$>$6.5 bits/symbol capacity, antenna diameters$>$0.7 m, yielding 22× throughput advantage for State-of-the-Art versus Low-Cost systems (550 vs 25 Gbps).

\section{Conclusion}
This paper has established a comprehensive theoretical framework for analyzing the fundamental performance limits of THz LEO ISL ISAC systems, moving beyond conventional AWGN-based models to incorporate the dominant physical realities of the space environment. Our unified signal model successfully integrates the intertwined effects of extreme kinematics, cascaded hardware impairments, and pointing error-dominated fading, providing a robust foundation for system analysis.

Our key findings reveal a paradigm shift from power-limited to hardware-limited design for THz ISLs. We derived a novel, closed-form expression for the communication capacity ceiling, demonstrating that performance saturates at a level determined by a quantifiable hardware quality factor $\Gamma_{\text{eff}}$ and phase noise. For sensing, we established the BCRLB for range and range-rate, quantifying the powerful quadratic scaling of accuracy with frequency, achieving sub-millimeter (1.06 µm) ranging precision with State-of-the-Art hardware at high SNR, ultimately floored by hardware-induced distortions. Crucially, our analysis of state-of-the-art components identified PA nonlinearity as the primary performance bottleneck, offering clear guidance for future hardware investment.

Through extensive numerical evaluation under the assumed hardware parameters, our analysis suggests that favorable operational conditions may exist in the sub-THz range (200-600 GHz) where the quadratic sensing gain with frequency is balanced against hardware quality degradation. This indicative finding provides initial guidance for frequency allocation in future THz ISL systems. The framework also yields a practical feasibility map for selecting antenna and PA configurations that can meet joint ISAC requirements. Future work should focus on the experimental validation of this framework and its extension to multi-node, networked ISAC for full 3D kinematic state estimation. The principles and limits established herein provide essential theoretical guidance for the architecture and design of next-generation 6G satellite constellations.

\bibliographystyle{IEEEtran}
\bibliography{references}

\begin{IEEEbiography}[{\includegraphics[width=1in,height=1.25in,clip]{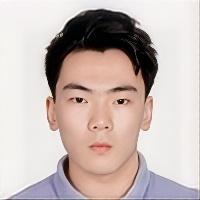}}]{Haofan Dong}
(hd489@cam.ac.uk) is a Ph.D. student in the Internet of Everything (IoE) Group, Department of Engineering, University of Cambridge, UK. He received his MRes from CEPS CDT based in UCL in 2023. His research interests include integrated sensing and communication (ISAC), space communications, and THz communications.
\end{IEEEbiography}

\begin{IEEEbiography}[{\includegraphics[width=1in,height=1.25in,clip]{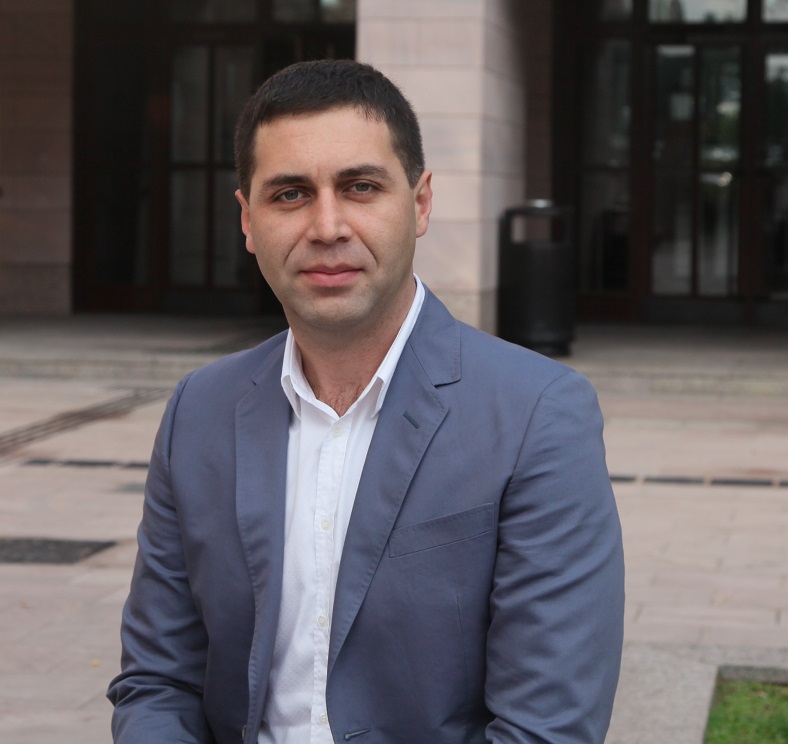}}]{Ozgur B. Akan}
(oba21@cam.ac.uk) received his Ph.D. degree from the School of Electrical and Computer Engineering, Georgia Institute of Technology, Atlanta, in 2004. He is currently the Head of the Internet of Everything (IoE) Group, Department of Engineering, University of Cambridge, and the Director of the Centre for NeXt-Generation Communications (CXC), Koç University. His research interests include wireless, nano-, and molecular communications, and the Internet of Everything.
\end{IEEEbiography}

\end{document}